# THE CCD CONTROLLERS AND DETECTORS

**Giovanni Bonano[1], Carlotta Bonoli[2], Fabio Bortoletto[2], Pietro Bruno[1], Maurizio Comari[3], Rosario Cosentino[1], Maurizio D'Alessandro[2], Daniela Fantine1[2], Enrico Giro[2], Salvatore Scuderi[1]**

[1].*Osservutorio Astrofisico di Catania*
[2].*Osservatorio Astronomico di Padova*
[3].*Osservawrio Astronomico di Trieste*

## Abstract

All the scientific instruments of the Italian National Telescope "Galileo" (TNG), as well as the tracking systems and the Shack-Hartmann wavefront systems use CCDs as detectors.
The CCD procurement is crucial to optimize the instruments performance, and, of course, equal importance assumes the design and realization of controllers able to drive various kinds of CCDs with a very low readout noise. Detectors characterization is of fundamental importance when they have to be used in scientific instrumentation. Various CCDs have been assembled in cryostats and tested at our laboratory in order to select the suitable detector for the optical instruments of the TNG. The relevant phases of the group activity are here described, as well as the commissioning at the telescope and the work that is in progress.

## 1. Introduction

It is well known that CCDs are the dominant detectors in the instrumentation for optical astronomy. The scientific instruments as well as the cameras for tracking and for the adaptive optics of the Italian National Telescope, apart from the infrared instrument, use CCDs as detectors.

A related activity from the CCD procurement to optimize the instruments performance to the design and realization of controllers is fundamental, as well as that of the detectors characterization. Various CCDs have been assembled in cryostats, tested at the detector laboratory of the Catania Astrophysical Observatory and installed at the TNG telescope during the commissioning. An activity of this type cannot finish with the commissioning, in fact improvements for the electronic controller and tests of a different type of detector are in progress.

## 2. Group activity

The activity of our group started a long time ago and during this very long period, the group has taken the responsibility for managing all things related to detectors from the electronic design of the CCD controller to the CCD selection and characterization. The people involved in this work are shown in Table I**.**

Table I. - People involved in this work

| NAME | INSTITUTE | ACTIVITY |
|---|---|---|
| Bonanno Giovanni | 0. A. Catania | detector & electronics |
| Bondi Carlotta | C. A. Padova | software |
| Bortoletto Fabio | 0. A. Padova | detector & electronics |
| Bruno Pietro | 0, A. Catania | software |
| Cali Antonio | 0. A. Catania | mechanics |
| Comari Maurizio | 0. A. Trieste | electronics |
| Cosentino Rosario | 0. A. Catania | detector & electronics |
| D'Alessandro Maurizio | 0. A. Padova | detector & electronics |
| Fantinel Daniela | 0. A. Padova | software |
| Farisato Giancarlo | 0. A. Padova | electronics |
| Giro Enrico | 0. A. Padova | software |
| Scuderi Salvatore | 0. A. Catania | detector & characterization |
| Timpanaro M. Cristina | 0. A. Catania | electronics |

The relevant items of the activity can be summarized as follows:

1. Electronic design of the CCD Controller:
2. Software design and computer interfacing;
3. Performance evaluation and electronic debugging of the various electronic boards and of the whole system;
4. CCD procurement for the TNG instruments: (JIG, LRS, SARG, Pointing and Tracking cameras;
5. Mosaic assembly and design of appropriate cold fingers;
6. Cooling systems and related vacuum equipment;
7. CCD characterization;
8. Commissioning at TNG and maintenance;
9. Research and development of future works.

### 3. The CCD controller

The first three activities have led us to develop a CCD controller that is able to drive all the currently available CCDs and that can be easily interfaced with a host computer which is linked to other workstations in a local area network. The controller is able to send and receive data/commands directly from the various workstations distributed on the telescope local area. The block diagram is shown in Figure l. It contains a Bus (the CCD controller bus) in which are plugged-in a sequencer board and at least one analogue board. The sequencer board generates the clock sequences. Each analogue board produces 8 programmable bias voltages, 16 clock drivers with independently programmable upper and lower levels and is also able to read and process four channels independently.

The controller is connected to a VME crate through optical fibers. The VME contains a shared memory for image storing, a transputer to send commands to the controller and an Ethernet connection to the TNG workstation system (see Figure 2).

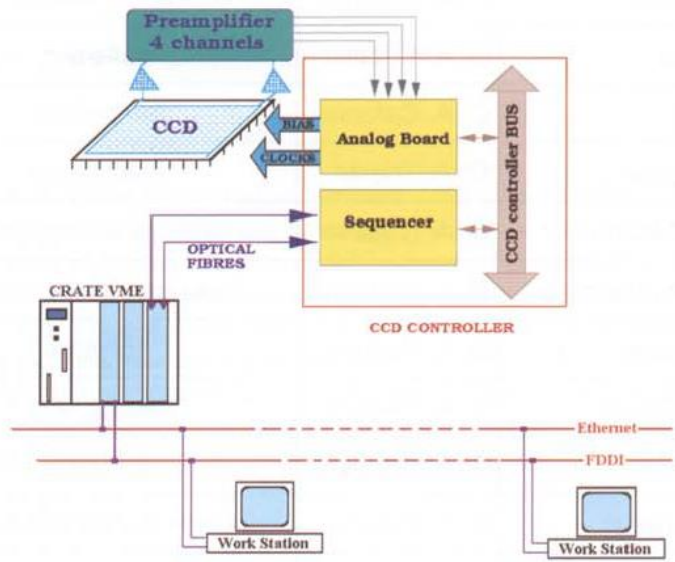

Figure 1. Block diagram or the CCD controller, connected to crate VME that is linked to a LAN.

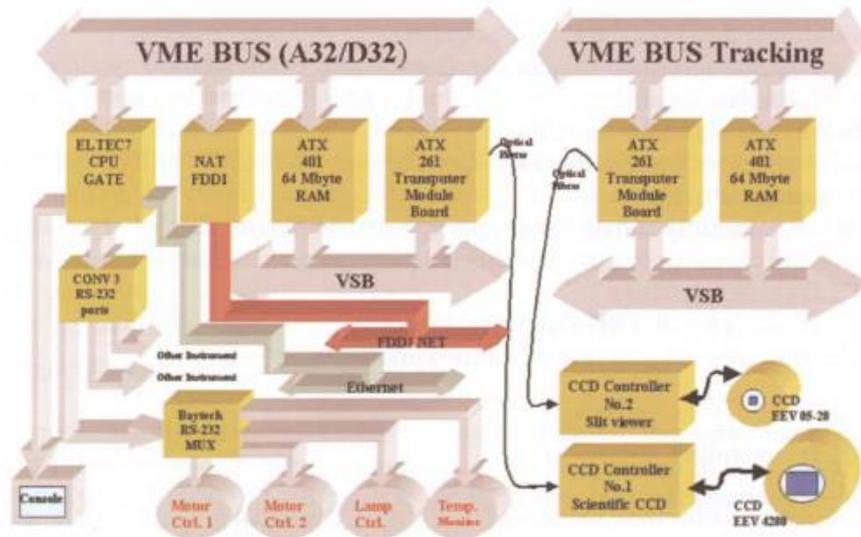

Figure 2. Block diagram of a typical crate VIVIE dedicated to an instrument. The crate VlivIE (host computer) provides interfacing with the CCD controllers.

The acquisition system allows the operator different readout modes (see Figure 3): frame transfer (tracking cameras), full frame (scientific cameras), binning on chip (1 x 2, 2 x 2, n x n) and readout of boxes (windowing).

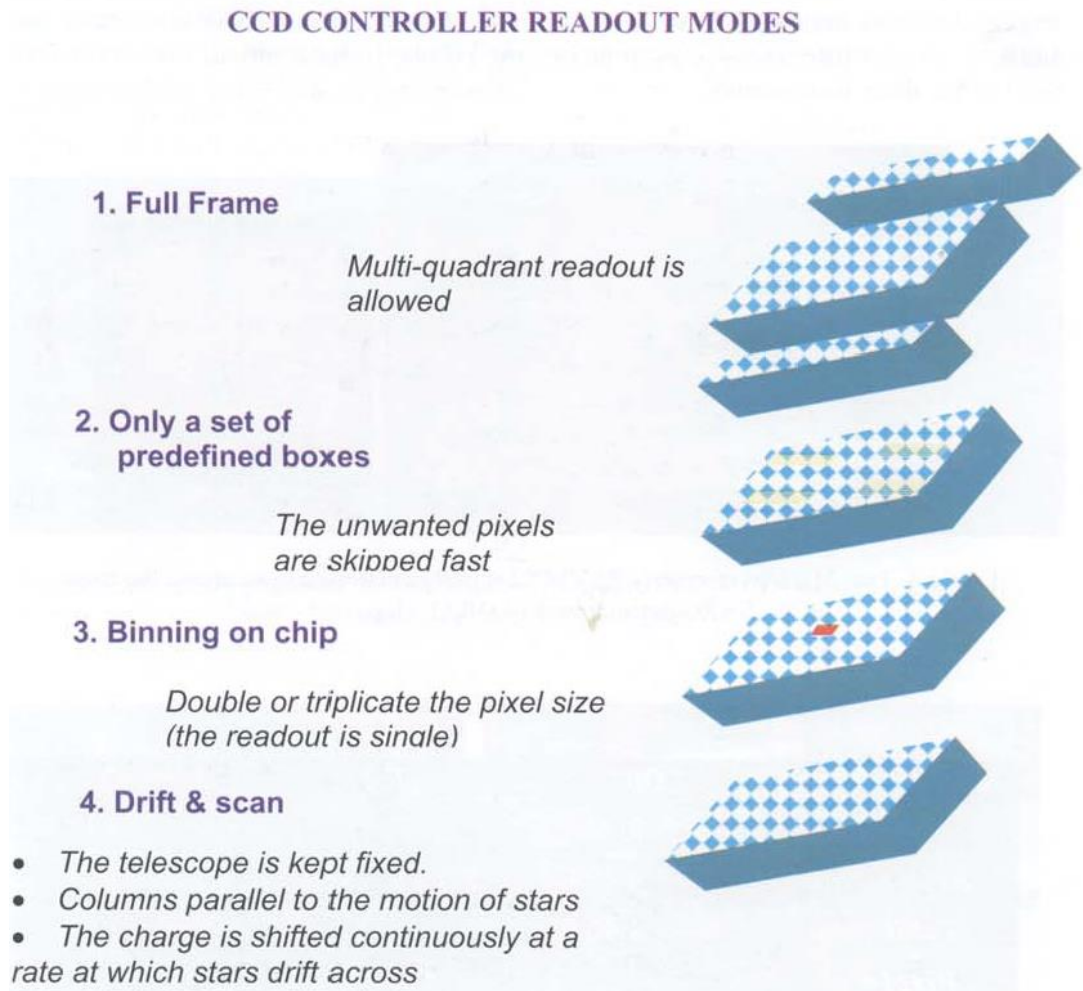

Figure 3. Typical Readout modes allowed by the CCD controller.

## 5. CCD procurement and performance evaluation

Performance evaluation has been made of current available CCDs in order to establish the best CCD detectors suitable for the TNG focal plane instruments. CCDs have to match the optical design and the scientific requirements. During this activity emphasis has been put on the spectral range, the achievable SIN and the uniformity of the response.

### 5.1 CCD detectors @ TNG

The CCD selection has given as result essentially two types of CCDs, one with a sensitive area of 2048x4096 pixels, manufactured by Marconi (formerly EEV) and the other with a sensitive area of 2048x2048 pixels, manufactured by LORAL and thinned and coated by the CCD laboratory of the Steward observatory (Michel Lesser). For both instruments, the optical imager Galileo (01G) and the high resolution spectrograph (SARG), two Marconi chips were assembled to form a total sensitive area of 4096x4096 pixels.

Figure 4 shows the two Marconi CCDs used to assemble the SARG detector and a LORAL chip, while Table 2 summarizes the relevant characteristics of the detectors used in the three instruments.

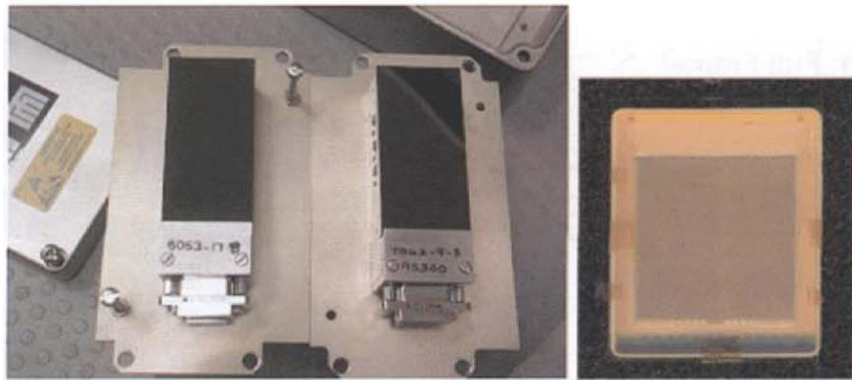

Figure 4. Two Marconi (formerly EEV) CCDs *(kfi panel)* used to assemble the mosaic of the SARG detector and a LORAL chip *(right panel)*.

Table 2. - Relevant characteristics of the detectors used in the three instruments.

| Characteristics | OIG | LRS | SARG |
|---|---|---|---|
| Detector | Mosaic of 2 CCD | 1 CCD | Mosaic of 2 CCD |
| Area single | 2048 X 4096 | 2048 X 2048 | 2048 X 4096 |
| Total area | 4096 X 4096 | - | 4096 X 4096 |
| Pixel size | 13.5 µm | 15 µm | 13.5 µm |
| Manufacturer | EEV | LORAL | EEV |
| Type | Thinned back ill. | Thinned back ill. | Thinned back ill. |
| MPP | NO | YES | NO |
| Working Temp. | -130 °C | -110 °C | -130 °C |
| UV Treatment | Ion Implant | Chemisorption | Ion Implant |
| AR Coating | YES | YES | |
| Grade | 2 - 3 | 2 | 2 - 2 |

## 6. Mosaic assembly and design of appropriate cold fingers

Both types of CCD are manufactured to allow the mounting of each CCD as close as possible to another to form a mosaic. We designed various cold plates to host the CCDs inside the cryostats and we arranged various tools to mount the mosaic on these plates. Figure 5 shows some mounting phases of the two Marconi CCDs used as detectors both for the OIG and for the SARG, while Figure 6 shows the assembled mosaic over the cold plate and the lower side of the cryostat.

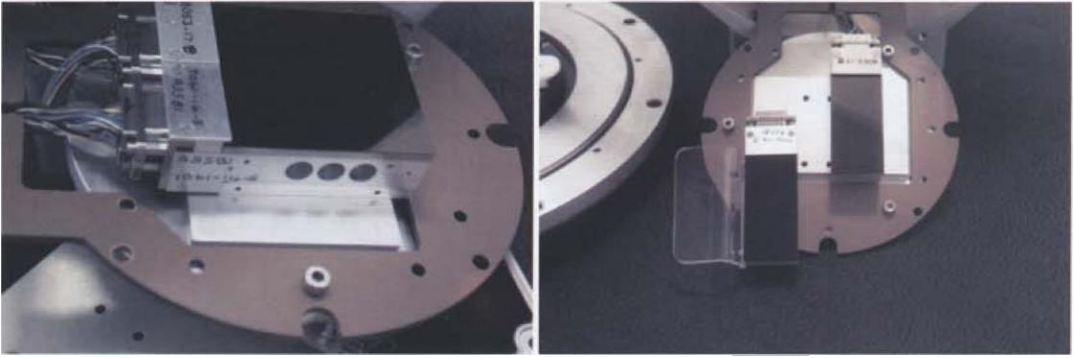

Figure 5. Mounting phases of the two Marconi (formerly EEV) CCDs used to assemble the mosaic of the SARG detector.

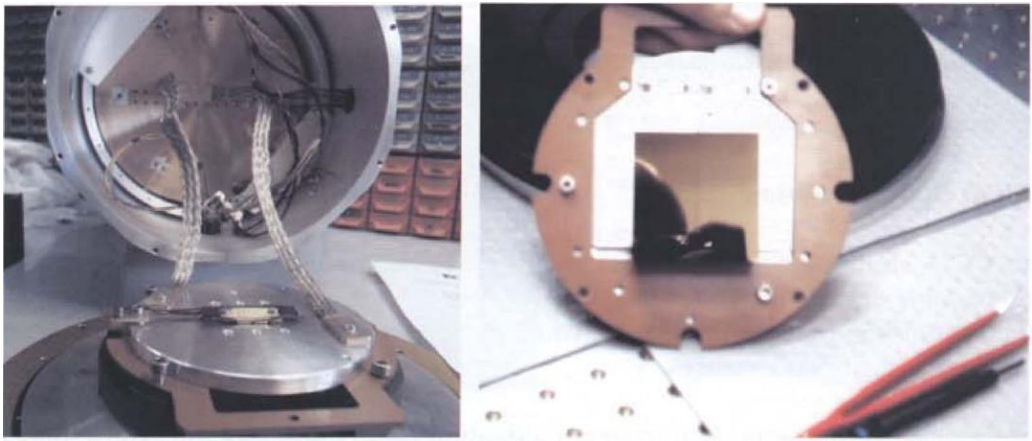

Figure 6. On the left the assembled mosaic over the cold plate. On the right, the flexible copper roads connected to the $LN_2$ plate are used to cool the plate holding the detector

As shown in Figure 6 the cooling is obtained through flexible roads and the working temperature is maintained stable by means of thermal resistors that act as heaters. A temperature controller controls the heating power of the resistors.
Finally the detector is mounted on an appropriate mechanical flange that includes the optical window and screwed on the cryostat as shown in Figure 7 (example of SARG detector).

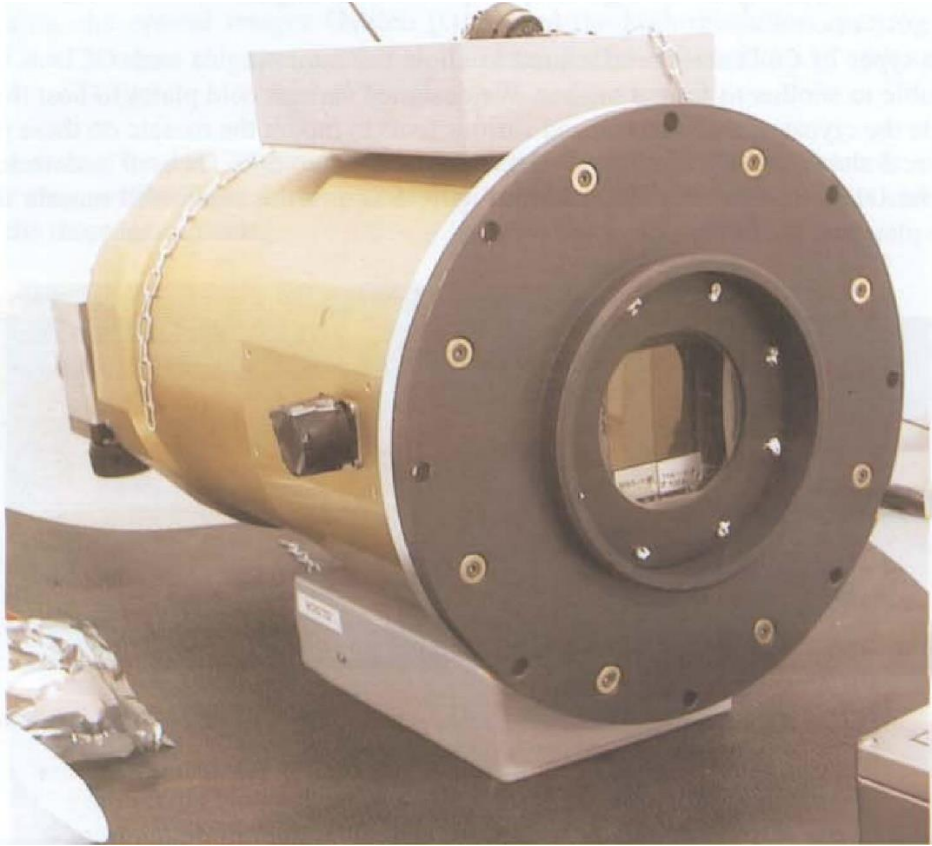

Figure 7. Cryostat complete with the mechanical flange that includes the optical window and the CCD mosaic.

## 7. CCD characterization

The CCD calibration activity was mainly carried out at Catania Astrophysical Observatory, while at the TNG we realized a small laboratory to tune-up the CCDs with the electronics in order to make them work at the telescope and to check the temporal behavior of their performances.

The Catania Astrophysical Observatory calibration facility allows a full electro-optical characterization of CCD detectors. The main parameters that can be measured are listed below:

- Gain ($e^-$/ADU);
- readout noise;
- charge transfer efficiency;
- linearity;
- quantum efficiency;
- dark current, hot pixel;
- cold pixel & traps.

The system gain, the readout noise and the CTE measurements are carried out using an Fe" x-ray source. Figure 8 shows the x ray camera than can be mechanically interfaced
120

with the cryostat thanks to a special flange realized on purpose. The cryostat window is removed to allow the x-rays impinging on the CCD.

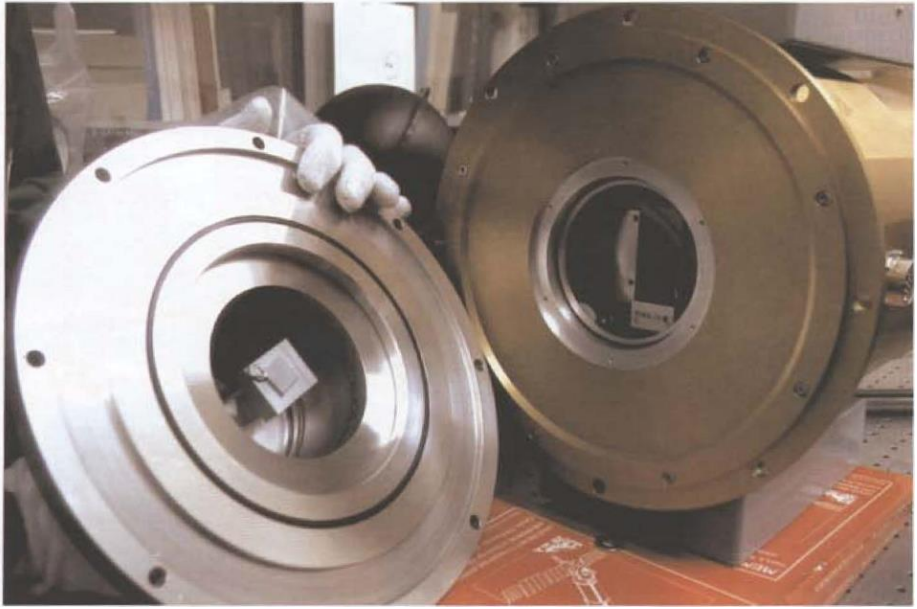

Figure 8. X-ray camera (source encapsulated on the small aluminum box to be interfaced with the Cryostat (without the window).

An example of gain and charge transfer efficiency (CTE) measurements is shown in Figure 9.

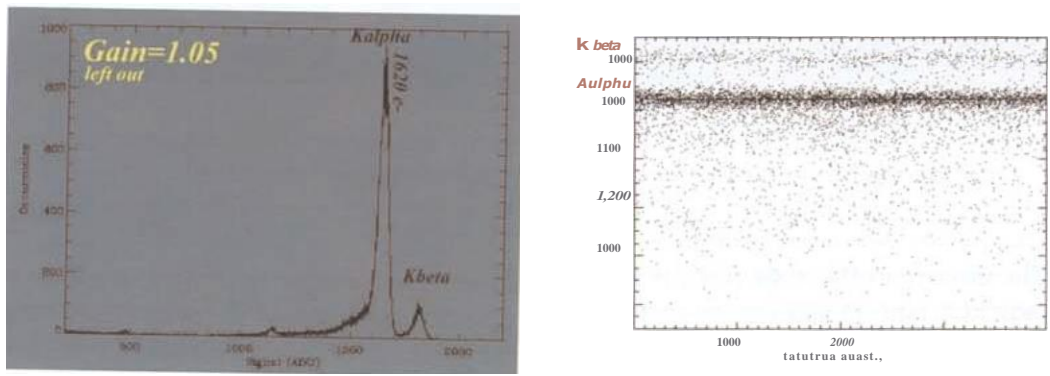

Figure 9. Example of gain (left panel) and CTE (right panel) measurement.

The main line (Kα) emitted by the source has an energy of 5.9 KeV and thus each photon can generate a charge of 1620 e$^-$ on a pixel. The measurement of the position of the line on the plot of Figure 9 (left panel) allows us to determine directly the system gain with a precision better than 1 %.

The CTE is measured by stacking together all the columns and by plotting the signal versus the number of pixel transfer (column number). The CTE is given by:

$$CTE = 1 - \frac{Charge\_loss}{1(1620e-)(Nt)}$$

where *Charge _loss* is the amount of charge lost in *Nt* transfers.

The readout noise is evaluated in the same image of the x-ray source by calculating the standard deviation in a box taken on an area without signal (overscan area).

For linearity and uniformity measurements we use a 20-inch integrating sphere optically connected with the QE measurement system through a quartz lens. The set-up apparatus is shown in Figure 10.

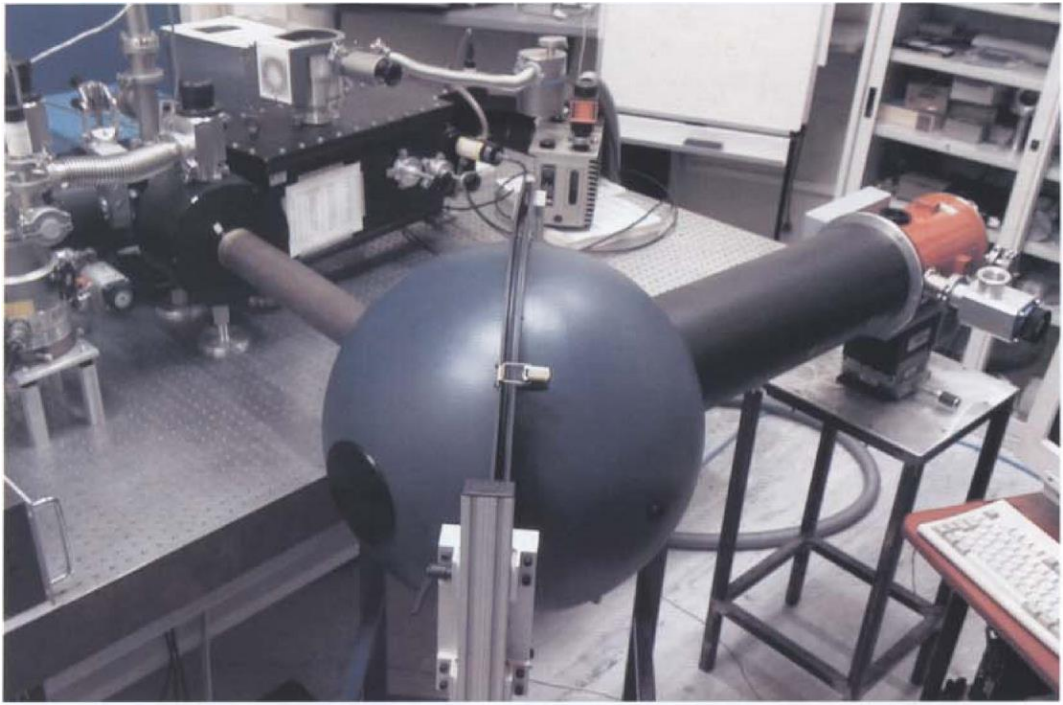

Figure 10. Set-up apparatus for linearity and uniformity measurements
at Catania Observatory Detector Laboratory

The quantum efficiency (QE) is measured illuminating the CCD with monochromatic radiation and comparing its response to that of a calibrated detector. To obtain QE measurements of the various TNG CCDs we used the instrumental apparatus shown in Figure 11, that is able to carry out QE measurements in the wavelength range 130 -1100 nm. The radiation emitted by two light sources (deuterium and xenon lamp) goes through a series of diaphragms and filters and is focussed onto the entrance slit of a monocromator. The dispersed radiation beam is thus divided by a beam splitter and focussed on a reference detector and on the CCD. A detailed description of this apparatus can be found in Bonanno et al 1996.

An example of QE measurements of two LORAL chips (the blue is that installed at LRS) is shown in Figure 12, and a plot of QE measurements of an EEV CCD used for the high resolution spectrograph SARG is shown in Figure 13.

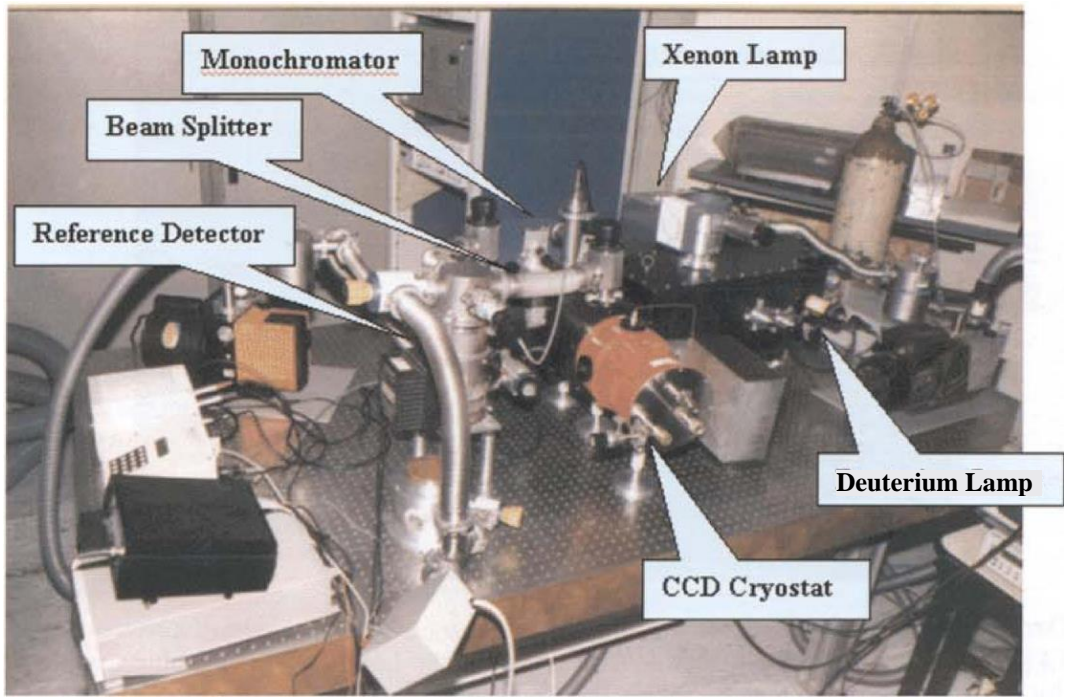

Figure 11. Set-up apparatus for quantum efficiency measurements in the same Laboratory, as Fig. 10

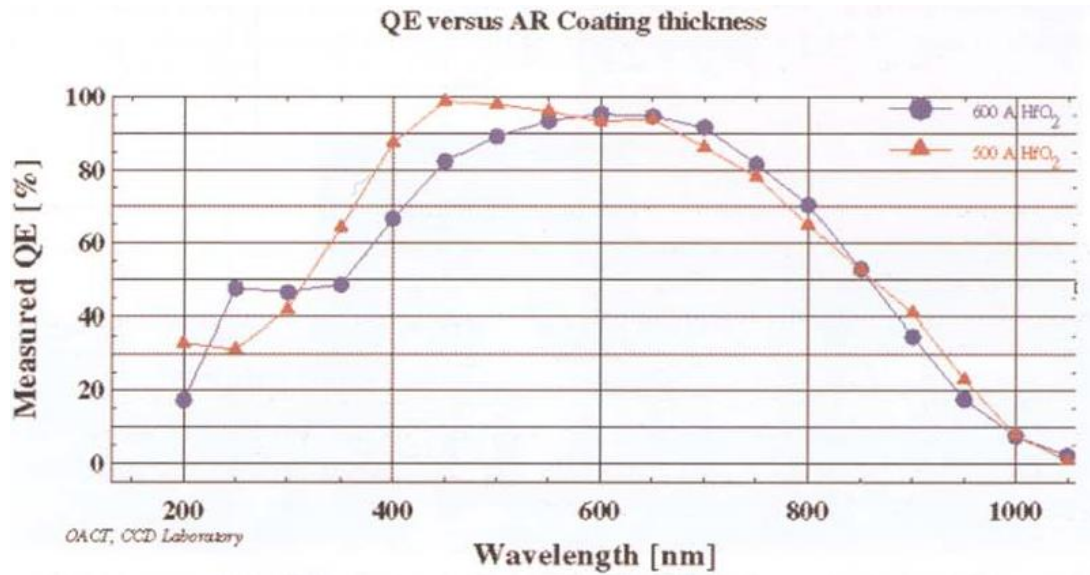

Figure 12. QE measurements of two LORAL chips (the blue one is that installed at LRS).

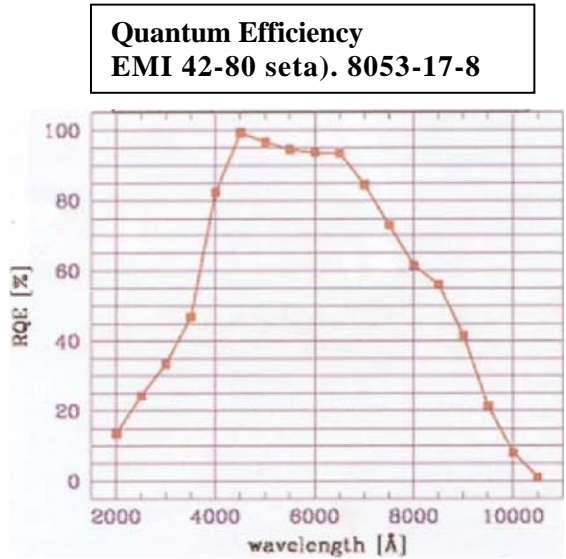

Figure 13. QE measurements of an EEV CCD used for the high resolution spectrograph SARG.

Dark current and pixel defects are also measured by taking long dark exposures, figure 14 shows on the left a 3600-second dark image of the Loral CCD (LRS chip), while on the right the generated map of hot pixels is presented.

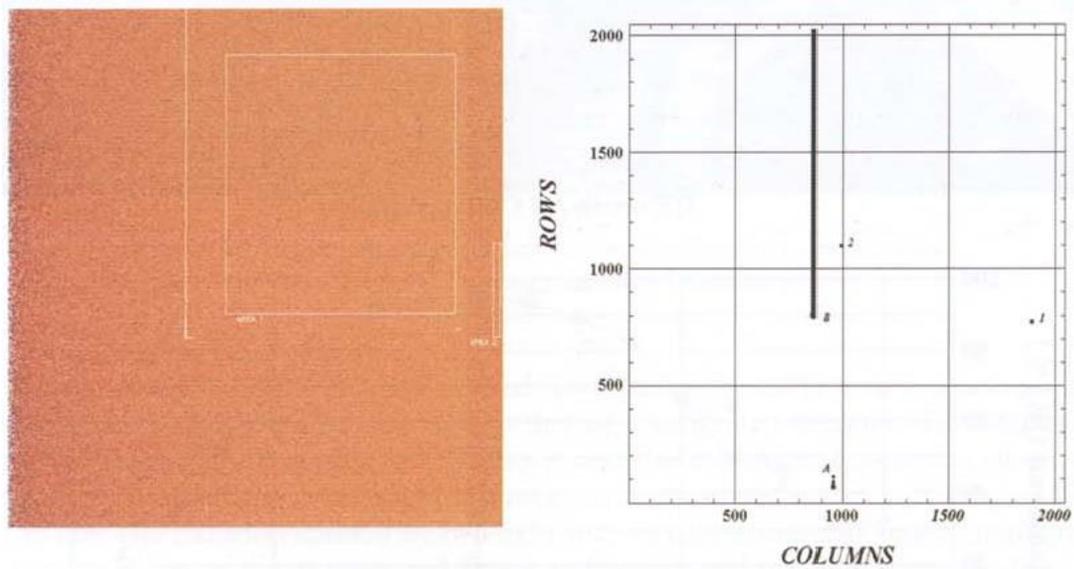

Figure 14. 3600-second dark image of the low resolution spectrograph CCD and the generated map of hot pixels.

The calibration facility at the TNG detectors laboratory consisting in a xenon lamp, two filter wheels with a series of interference filters and a series of neutral Filters, a 20-inch integrating sphere and a reference photodiode, is shown in Figure 15.

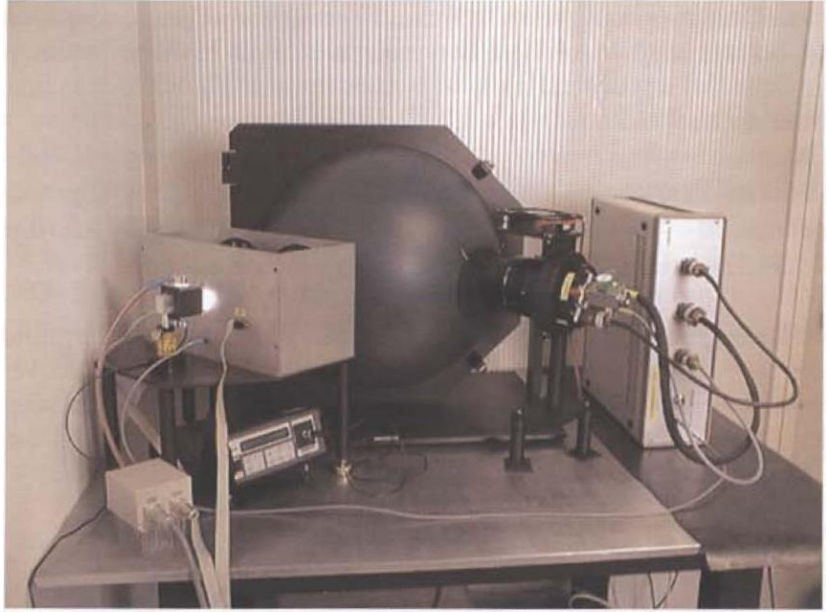

Figure 15. The calibration facility at the TNG detectors laboratory.

## 8. Commissioning at the TNG

The group activity was pinned by working jointly with CGI group, and the definitive installation of the various instruments at the telescope has been done. During the commissioning a lot of work related to the detectors has been done from the vacuum operation to the cooling of the cryostats, to mounting at the telescope. Figure 16 shows some cryostats under vacuum operation, the OIG mounted at the "A" rotator adapter, and the LRS already mounted while SARG is being mounted at the "13" rotator adapter.

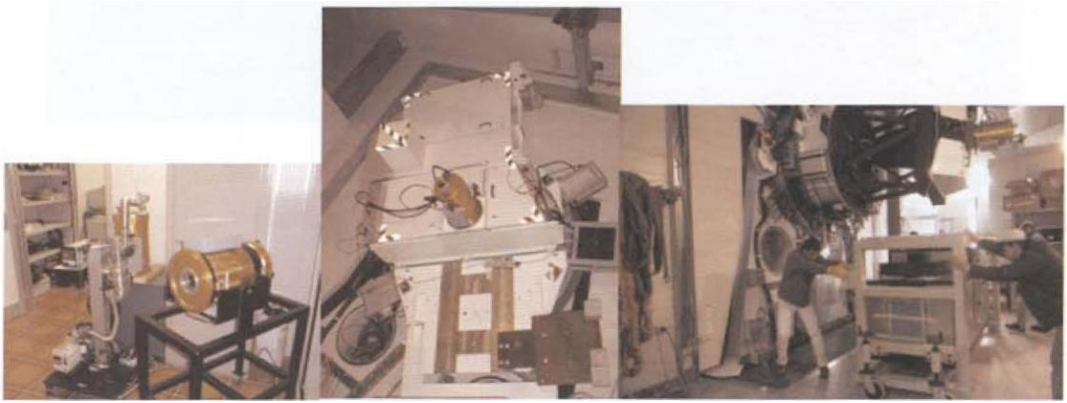

Figure 16. Some phases of the commissioning at TNG

## 9. Work in progress

The commissioning at the telescope is not the final step for our group, as usual, because two projects have already been started and are **in** progress.

One is devoted to improving the OIG camera, and the other to improve the CCD controller. A brief description of these two projects is given in the following paragraphs.

### 9.1. New CCD for the 016

To improve the OIG camera we selected a 2048 x 2048 pixel scientific grade CCD. This CCD, manufactured by SITe, is characterized by a pixel size of 24 µm, thus having a 5x5 cm$^2$ of sensitive area without inconvenient gaps on the image.

The project is very ambitious because an X-Y movement system that holds the cold plate is assembled inside the cryostat. This system allows the CCD to move quickly in both directions and improves the image quality taking advantage of the adaptive optics in use at the TNG. Figure 17 shows the CCD mounted on the board and the cold plate that accommodates the detector.

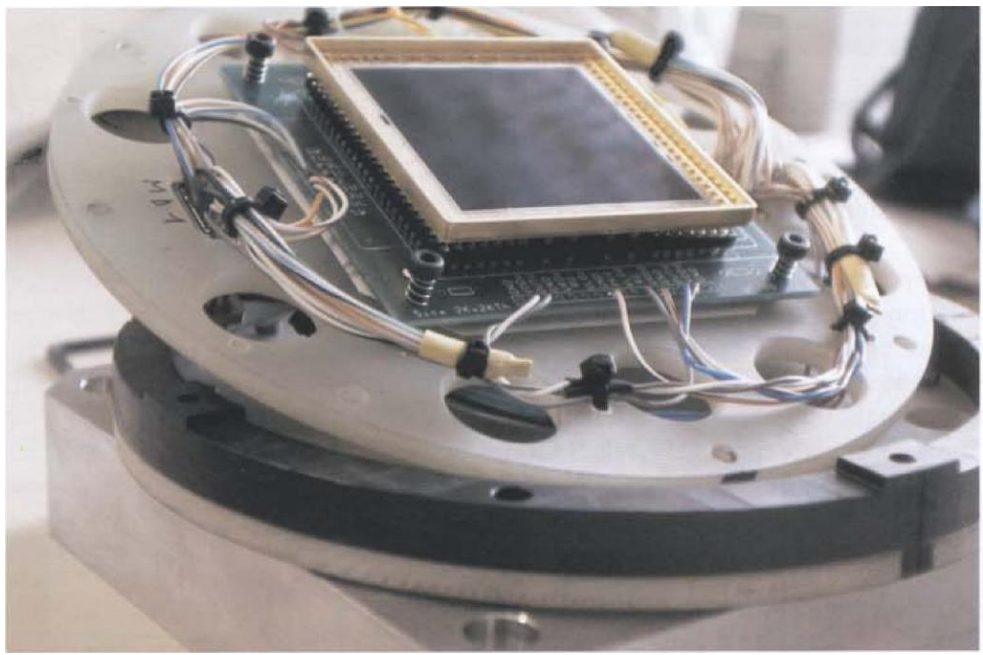

Figure 17. SITe 2048 x 2048 pixel CCD (24 µm pixel size) mounted on the cold plate.

### 9.2. New generation CCD controller for the TNG

The idea to have at the TNG a new CCD controller, based on new electronic components, able to easily drive mosaics 5-10 times faster than the previous CCD controller, began a few years ago, when we observed that to read the two EE V chips (16 Megapixels) of the OIG, the CCD controller spent about 250 seconds to read the whole image. This time, of course, is not adequate to the current readout allowed by new CCDs.

A concrete project has now started. The compactness, the weight, the readout speed and the noise performances are the key points that drove the design. A lot of electrical schemes have been designed, and finally a new architecture has been achieved. All digital sequences for the CCDs, are generated by an interface located on the host computer, and sent to the CCD electronics through a high speed full duplex fiber link. The signals are replicated on the SEQUENCER board that provides also the CCD clock signals. In fact the board has digital-to-analogue converters for the upper and lower level voltages of the clocks. We adopted the PCI bus standard to interface the CCD controller with the host computer. The AMCC 55933 PCI matchmaker controller chip is used for this purpose. To handle the incoming data we use a FIFO memory and the AMCC 55933 chip. The CCD controller, excluding the power supply, is constituted essentially by three electronic boards: two double euro-cards and one extended PCI board. Figure 18 shows the block diagram of all the architecture.

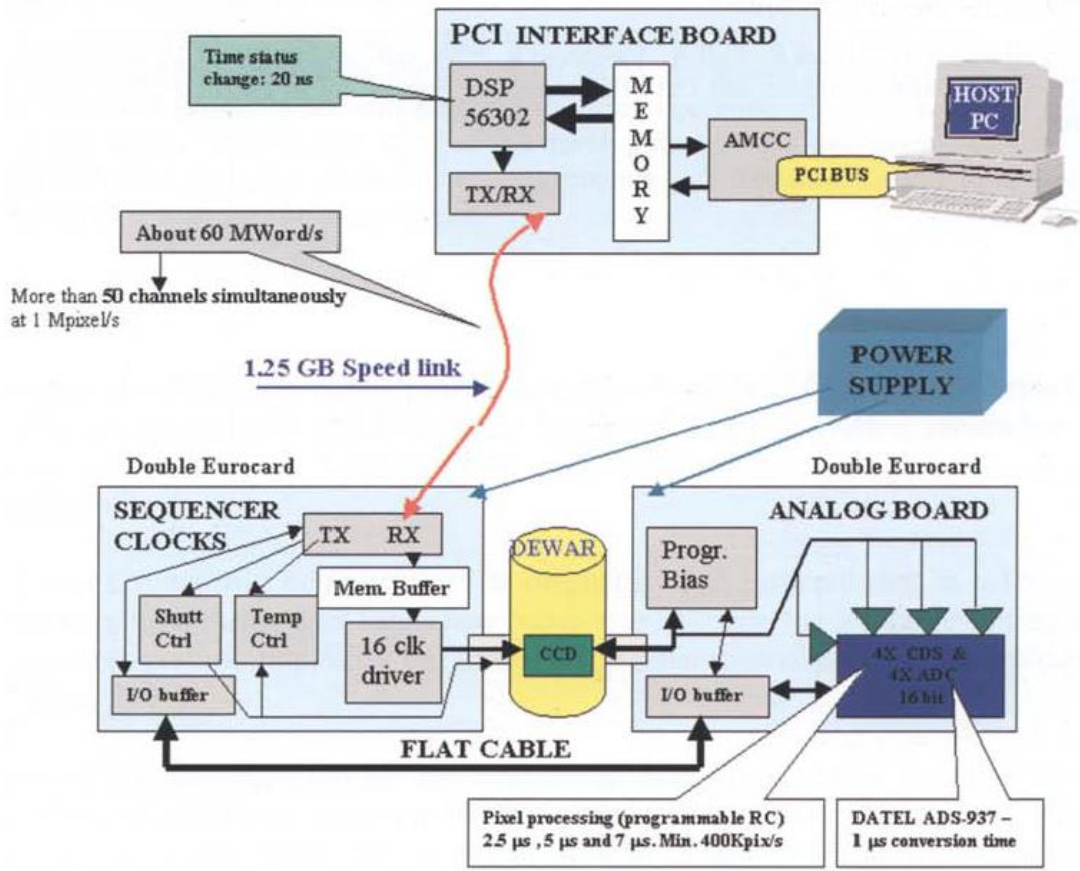

Figure 18. Block diagram of all the architecture of the new generation CCD controller for the TNG.

This architecture allows us to drive a mosaic of four CCDs with one output or a mosaic of two CCDs with two outputs, by using just two boxes mounted directly on the CCD cryostat. Only two fibers link the CCD controller to the host computer, one to send the digital sequences and the other to receive the data pixel. A flat cable connects the analogue box to the sequencer box to send the acquired and converted pixel signal.

With a Correlated Double Sampler (CDS) of 2.5 µs at minimum, this new CCD controller will allow an acquisition rate of 400 k pxl/s per channel, and thanks to the high speed link, many channels will be read at a time, which is very useful for large mosaics.